\documentclass[fleqn,usenatbib]{mnras}

\usepackage{newtxtext,newtxmath}

\usepackage[T1]{fontenc}

\DeclareRobustCommand{\VAN}[3]{#2}
\let\VANthebibliography\thebibliography
\def\thebibliography{\DeclareRobustCommand{\VAN}[3]{##3}\VANthebibliography}

\usepackage{graphicx}	% Including figure files
\usepackage{amsmath}	% Advanced maths commands
\usepackage{amssymb}	% Extra maths symbols
\usepackage{ulem}

\usepackage{color}

\newcommand{\mph}{$h^{-1}\rm Mpc\,$}
\newcommand{\msol}{$\rm M_{\odot}\,$}
\newcommand{\msolh}{$h^{-1}\rm M_{\odot}\,$}
\newcommand{\rvir}{$\rm r_{\rm vir}$}
\title[Void evolution imprints on halo baryon processes]{Imprints of the cosmic void evolution on the baryon processes inside galaxy haloes}

\author[Rodr\'iguez M. et al.]
{
Agustín M. Rodr\'iguez Medrano,$^{1,2}$\thanks{E-mail: arodriguez@unc.edu.ar}
Dante J. Paz,$^{1,3}$
Federico A. Stasyszyn$^{1,3}$ and
\newauthor
Andrés N. Ruiz$^{1,3}$ 
\\
$^{1}$Instituto de Astronom\'ia Te\'orica y Experimental, CONICET-UNC, Laprida 854, X5000BGR, C\'ordoba, Argentina\\
$^{2}$Facultad de Matem\'atica, Astronom\'ia, F\'isica y Computación, UNC, Av. Medina Allende s/n, Ciudad Universitaria, X5000HUA, Córdoba, Argentina\\
$^{3}$Observatorio Astronómico de Córdoba, UNC, Laprida 854, X5000BGR, C\'ordoba, Argentina\\
}

\date{Accepted XXX. Received YYY; in original form ZZZ}

\pubyear{2021}

\begin{document}
\label{firstpage}
\pagerange{\pageref{firstpage}--\pageref{lastpage}}
\maketitle

\begin{abstract}

Cosmic voids provide a unique environment to study galaxy formation and evolution. 
In this paper, we analyse a set of hydrodynamic zoom-in simulations of voids, to analyse in detail their inner structures. 
These voids were identified in a cosmological simulation and classified according to their surrounding dynamics at very large scales: whether they are in expansion or contraction at their outskirts. 
We study how these environments and the dynamics of voids impact the baryonic processes inside haloes and their mechanisms of formation and evolution. 
We find an under-abundance of processed gas within the voids that can be associated with the lack of massive haloes. 
By studying the dynamical phase-space diagram of haloes and the halo-particle correlation function, we find that haloes inside of contracting voids are slightly affected by the presence of bigger structures, in comparison to haloes in the inner regions of expanding voids. 
Consistent signals are obtained both when using dark matter and gas particles.
We show that the halo assembly depends on the void dynamical state: haloes in expanding voids assemble slowly than those in contracting voids and in the general universe. %
This difference in the assembly impacts the baryonic evolution of haloes. 
Overall the redshift range analysed, haloes in voids have less baryon content than haloes in the general universe and particularly at $z=0$ less stellar content.
Our results suggest that the large scale void environment modulate the baryonic process occurring inside haloes according to the void dynamical state.

\end{abstract}
\begin{keywords}
large-scale structure of Universe -- galaxies: haloes -- galaxies: formation -- galaxies: evolution -- methods: numerical
\end{keywords}

%%%%%%%%%%%%%%%%%%%%%%%%%%%%%%%%%%%%%%%%%%%%%%%%%%

%%%%%%%%%%%%%%%%% BODY OF PAPER %%%%%%%%%%%%%%%%%%

\section{Introduction}
\label{sec:intro}

With the advent of large volume galaxy surveys \citep[e.g. :][]{Colless2003,Tegmark2004, Huchra2005} and cosmological simulations \citep{Davis1985,White1987,Dolag2006,Cautun2014}, numerous studies have revealed how galaxies populate the Universe, forming a cosmic web of filaments, walls, voids and nodes. Cosmological voids are the most underdense regions of the Universe and then one of the most extreme environments where galaxies can form and evolve. These regions may contain a few galaxies or even lack of them and their properties are very dependent on the void definitions used, nevertheless, there is a consensus that they constitute most of the volume of the Universe \citep{Pan2012}. 

Numerical simulation studies have made possible to characterise both the content of the voids and their dynamics. The inner region of these subdense regions are expanding \citep{Sheth2004}, making interactions between galaxies in their interiors infrequent \citep[][ and references therein]{VanDeWeygaert2011}. Therefore, galaxies in the interiors of the voids can be considered more pristine, being less affected by ram-pressure stripping or harassment processes. This has been studied with different approaches in simulations, founding that void haloes assemble later than haloes in denser environments  \citep[][ among others]{Tonnensen2015,Tojeiro2017,Martizzi2020,Alfaro2020,Zhang2021}.
However, the evolution of voids does not prevent the formation of tenuous networks of filaments in their interiors \citep{Gottlobler2003, Rieder2013, Beygu2013, Alpaslan2014}.

Observational studies indicate that void galaxies tend to be blue, faint, with late morphologies and high star formation \citep{Rojas2004, Rojas2005, Hoyle2005, Patiri2006, Ceccarelli2008, vonBenda2008, Hoyle2012, Kreckel2014, Ricciardelli2017, Florez2021}. 
This is in part due to a large amount of low-mass galaxies in these regions, but it is a source of debate if this also reflects an environmental influence. \citet{Ricciardelli2014} using data from the Sloan Digital Sky Survey (SDSS) DR7 found that, at fixed stellar mass, the proportion of star-forming galaxies in voids is higher than in a control sample. This result can be used to argue that there is an influence of the environment on the modulation of star formation in galaxies. In contraposition, \citet{Douglass2019} did not find evidence of the influence of the void environment in astrophysical properties of void galaxies. Using observational data from the SDSS MaNGA DR15, they found that the stellar-to-halo-mass relation is independent of the environment.

Beyond astrophysical properties, they also present differences associated with clustering. \citet{Ruiz2019} show a low amplitude of the galaxy-galaxy correlation function for galaxies inside of voids. \citet{Alfaro2020} associate this with a difference in the Halo Occupation Distribution (HOD) that describes the probability distribution that haloes contain a certain number of galaxies. Their results indicate that haloes in voids contain a minor number of galaxies than haloes in the general universe. 

The population of voids can be divided according to the dynamics of the void surroundings.
This introduces classification of voids made by \citet{Sheth2004}, where they divided voids into two categories: \textit{void-in-void} and \textit{void-in-cloud}. The first type of voids corresponds to those embedded in an underdense environment with expanding walls while the second are voids that are embedded in an overdense environment and their walls in contraction.

This dichotomy was firstly detected in observations by \citet{Ceccarelli2013}, where they classify the voids as R-type or S-type according to its integrated density contrast profile. R-type voids are those with a \textit{rising} profile and with expanding walls. The integrated density profile is always below the mean density of the universe reaching only this value at large distances, then it is said the void is compensated. S-type voids have a \textit{shell} in their density profile and walls in contraction. 
The integrated density profile
exceeds the mean density of the Universe at the void surroundings,
thus it is defined as an overcompensated region, to then
reach the mean density at larger scales
This dynamic classification also correlates with voids sizes. Small voids are typically surrounded by overdense walls. In contrast, large voids typically have subdense environments and continuously increasing integrated density profiles. Thus, the fraction of S-type voids decreases as larger voids are considered.  

Although the characteristics of galaxies in these environments have been studied in-depth, relatively few studies have been conducted on whether the large-scale environment of voids and their dynamics affects the properties of embedded galaxies.
However, \citet{Ruiz2015} found that the nonlinear dynamics of the haloes in the internal regions of the voids depends on their type.
Motivated by these results, in this work we intend to study how this dichotomy in the properties of voids affects the formation, evolution and characteristics of haloes and the astrophysical processes within them.

This paper is organised as follows. In Sec. \ref{sec:simu} we describe the simulations used in this work and we present the codes used to identified voids, haloes and mergers trees. In Sec. \ref{sec:cosm_prop} we characterise the void environment and its surroundings paying special attention to the properties of the gas and its different stages. In Sec. \ref{sec:DM_haloes} we analyse the halo properties studying their phase-spaces in voids, his near regions and the general Universe, always separating according to the voids-large scale environment. We also study their evolution with redshift. Finally. in Sec. \ref{sec:summary} we present a summary and our conclusion of this work.  

\section{Simulations and numerical methods}
\label{sec:simu}
  
As we mentioned at the end of the previous section, in this work we are interested in studying the internal structure of voids based on their dynamics and the evolution of their environment on large scales.
In particular, considering the dynamic classification of voids of \citet{Ceccarelli2013} mentioned above.
Therefore, we adopted a strategy to run a base cosmological simulation to model the larger-scale modes responsible for this dichotomy in the void evolution, and then we run high-resolution hydro-resimulations of selected void regions to model the inner and near structure of voids.

\subsection{Cosmological simulation and voids identification}
\label{subsec:cosmosim}

As the cosmological base for void identification, we run a dark matter only simulation with a cubic volume of $500$\mph on side and $512^{3}$ particles. The cosmological parameter used are corresponding to a $\Lambda$ cold dark matter Universe and are $\sigma_{8}=0.8$, 
$\Omega_{\rm m}=0.3, \Omega_{\Lambda}=0.7, \Omega_{\rm bar}=0.04$ and $h=0.7$.  The initial condition was generated using the public code {\sc music} \citep{MUSIC} and the integration was made with the {\sc gizmo} code \citep{GIZMO2015}. The haloes in this simulation were identified using the {\sc rockstar} code \citep{Rockstar}.
The virial mass $\rm M_{vir}$ is defined as the mass that encloses a sphere of radius $\rm r_{vir}$ where the mean density is 200 times the critical density of the Universe.

In order to define void regions, we use the spherical void finder algorithm described in \citet{Ruiz2015}, which is a modified version of the algorithm presented by \citet{Padilla2005}. This algorithm computes the largest sphere in an underdense region with an integrated density contrast ($\Delta$) below a certain threshold,

\begin{equation}
\Delta(R_{\rm void})=\frac{3}{R_{\rm void}^{3}}\int_{0}^{R_{\rm void}} \delta(r)r^{2}dr<-0.9
\end{equation}
where the differential density contrast $\delta$ is the contrast over
the mean density of haloes.
The algorithm avoid overlapping among the spheres. 

The voids were identified using haloes as tracers with a minimum mass of $20$ particles ($491367$ objects), that is, with a threshold mass of $1.4\times10^{12}$\msolh.
The void catalogue generated contains $1384$ objects with a radius in the range of $\sim9-30$ \mph. We calculate the integrated density profile for each void and split the samples in R-type (voids in expansion at all scales) or S-type (voids shrinking at large scales) according to the criteria of \cite{Ceccarelli2013}.  We identify 711 S-type voids and 673 R-type voids. About $60\%$ of the small voids ($\sim10$ \mph) are classified as type S, while at the large size end ($\sim25$\mph Mpc) this fraction reaches $25\%$.

We selected $7$ voids in the dark matter only simulation with radii between $9-10$ \mph of which 3 voids are of type S with a very well defined overcompensation behaviour in density and 4 are R-type voids with a clear rising density profile.
The sizes of the selected voids were chosen to take into account a tradeoff between obtaining a physically meaningful void region (that is, away from the shot noise regime and with well defined dynamics) and the available resolution of the hydrodynamic simulations given the computational resources.

\subsection{Hydrodynamic zoom-in simulations of voids and the cosmological reference simulation}
\label{subsec:hidrosim}
\begin{figure}
    \centering
    \includegraphics[width=\columnwidth]{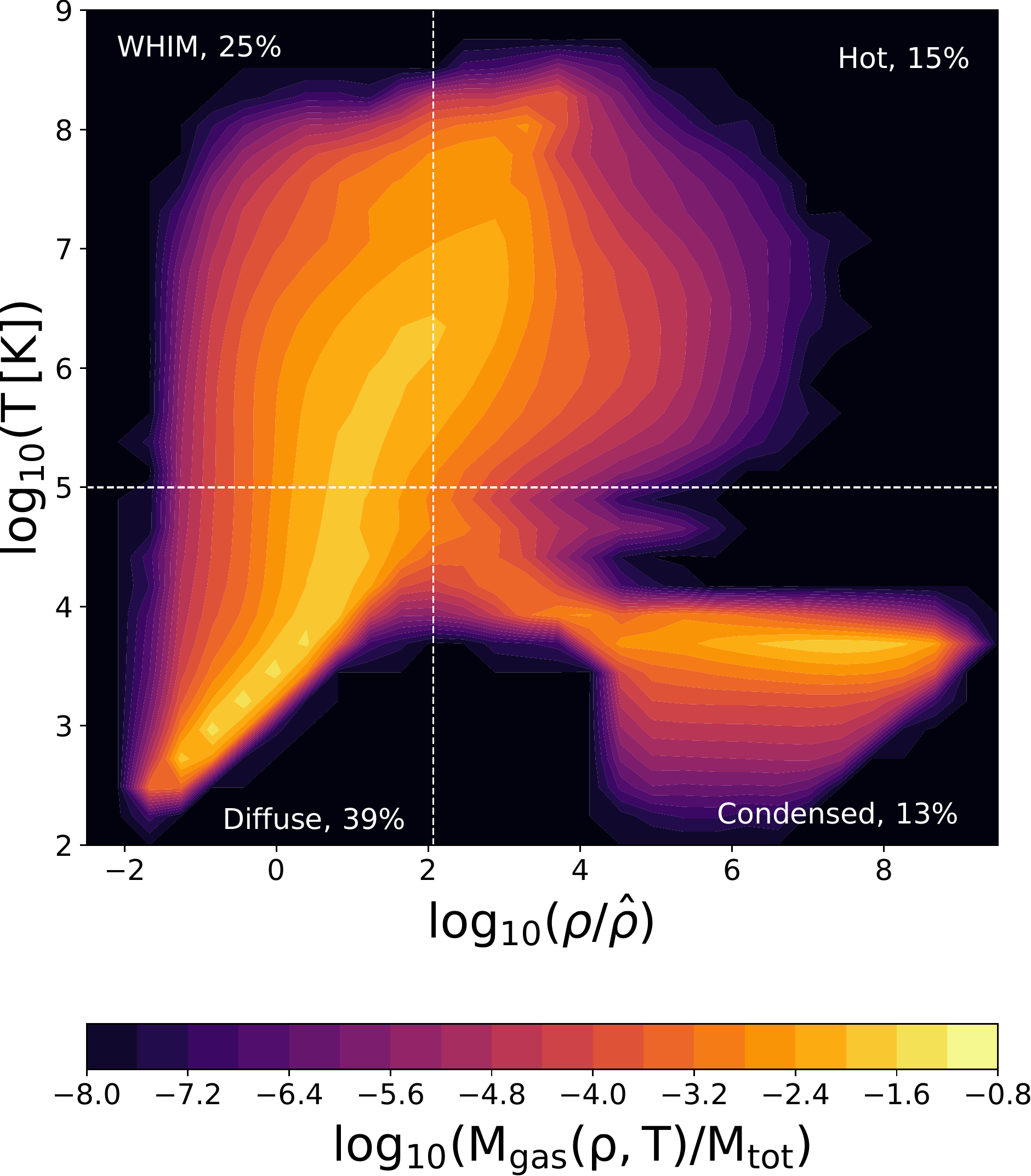}
    \caption{ Density-temperature phase diagram from the reference simulation box at $z=0$. Densities are normalised by mean baryon density $\hat{\rho}$. Black lines indicate the thresholds used to divide the gas into four phases, following \citet{Huang2019}. The colour map indicates the fraction of gas mass in each bin with respect to the total baryonic mass. For each phase, we indicate the percentage of gas relative to the baryons. 
    }
    \label{DF_ref}

\end{figure}

\begin{figure*}
    \centering
    \includegraphics[width=16cm]{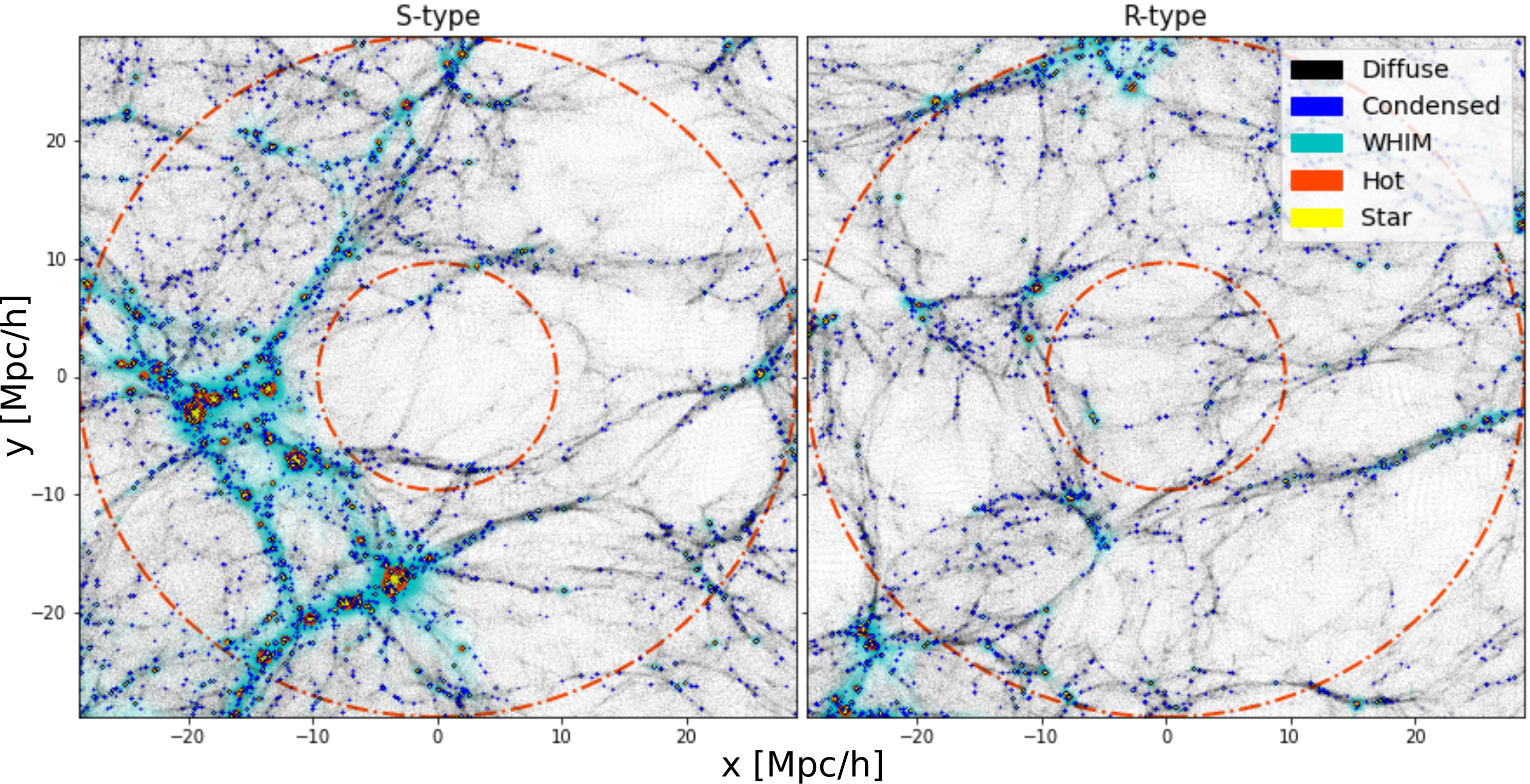}
    \caption{Slices of gas particle distributions for two void resimulations. On the left panel, there is an example of an S-type void while on the right panel there is an R-type void. Both regions have a radius, $r_{\rm void}$, close to $\sim10$\mph, indicated by the inner circle. The external circle enclose a region of $3\,r_{\rm void}$. Different colours indicate particles of different baryon stages: diffuse gas in black, condensed gas in blue, WHIM gas in green, hot gas in red and stellar particles in yellow. }
    \label{slides}
\end{figure*}

The hydrodynamic simulations in this work were performed using the classical implementation of sub-grid modelling of cooling, star formation and thermal feedback from \citet{Springel2003} with the {\sc gizmo} code \citep{GIZMO2015}. We decide to use this classic sub-grid model for star formation to have results very well established about the general large scale structure, and comparable with several previous works of the community \citep{Moscardini2004,Springel2004, Madau2014}. We decided not to use a cutting-edge model to avoid the complication of identifying whether potential differences arise from the modelling itself or from the peculiarities of the void cosmological environment.

In order to perform the hydro-resimulations of the selected void sample, we identify all the particles at $z=0$ in a sphere centred in the void with integrated density contrast $\Delta(r)=0$. In this way, these regions are selected so that they do not expand or contract and, therefore, we avoid the effects of contamination with low tidal resolution particles in the resimulation runs.
The particles contained in these regions are traced back to the DM-only initial condition, thus defining the domain to be resimulated.
Using the {\sc music} code \citep{MUSIC}, we introduce in these domain dark matter and gas particles with high resolution, with masses of $9.2\times10^{8}$\msolh and $1.8\times10^{8}$\msolh, respectively. The zoom-in boxes are in the range of $60-110$ \mph per side. In the extended zone, we introduce three levels of resolution to avoid contamination effects in the high-resolution domain. The simulation parameters for each void are summarised in Table \ref{tab:resims}. 
Although the void radius, $\rm r_{void}$, is similar in all cases the resimulated volumes span a large interval. This is due to the particularities of each void make it necessary to take a size large enough to avoid mixing of high and low resolution particles.
This issue made some of these resimulations numerically expensive, however, we try to re-simulate the biggest possible box enclosing the scale where the voids were compensated, which in general correspond to a scale between $\sim 3-4\, {\rm r_{void}}$.

\begin{table*}
\caption{Summary information on the characteristics of the simulations used in this work. The 7 resimulations were carried out from the base-simulation. The zoom-in region is indicated by the resimulation size. $\rm M_{DM}$ is the mass of the high resolution dark matter particle and $\rm M_{gas}$ is the mass of the gas particle. $\rm N_{gas}$ indicate the number of gas particles at the beginning of the simulation and $\rm N_{part\,tot}$ is the total number of particles used, including gas particles, high and low resolution dark matter particles. The void radius is $\rm r_{void}$. The Reference simulation is a periodic box with the same resolution as void resimulations.}

\centering
\begin{tabular}{ccccc}
\hline \noalign{}
Label & Box size & $M_{\rm DM}$ & $M_{\rm gas}$ & $N_{\rm part}$ \\
      & \mph     &  \msolh       & \msolh         &   \\
\hline

Base  & 500      & $ 7.3\times10^{10}$ & -                 & $512^{3}$  \\ 
Reference &  125 & $ 9.2\times10^{8}$ & $1.8\times10^{8}$ & $512^{3}$  \\

\hline

\end{tabular}

\begin{tabular}{ccccccccc}
\hline \noalign{}

Label & resim. size-x & resim. size-y & resim. size-z  & $M_{\rm DM}$ & $M_{\rm Gas}$ & $N_{\rm Tot}$ & $N_{\rm Gas}$ & $r_{\rm void}$ \\
      & \mph & \mph & \mph  & \msolh & \msolh & $\sim 10^{6}$ & $\sim 10^{6}$ & \mph \\

\hline

R1 & $57.65$  & $67.09$  & $80.54$ & $ 9.2\times10^{8}$ & $1.8\times10^{8}$ & $90$ & $36$ & 9.59 \\ 
R2 & $101.43$ & $112.03$ & $97.64$ & $ 9.2\times10^{8}$ & $1.8\times10^{8}$ & $213$ &  $97$ & 9.55 \\
R3 & $114.77$ & $107.04$ & $114.71$ & $ 9.2\times10^{8}$ & $1.8\times10^{8}$ & $150$ &  $66$ & 9.62 \\
R4 & $78.94$ & $71.16$ & $70.25$ & $ 9.2\times10^{8}$ & $1.8\times10^{8}$ & $86$ &  $34$ & 9.39 \\
S1 & $69.99$ & $72.93$ & $ 75.63$ & $ 9.2\times10^{8}$ & $1.8\times10^{8}$ & $77$ &  $30$ & 9.75 \\
S2 & $81.67$ & $75.11$ & $87.37$ & $ 9.2\times10^{8}$ & $1.8\times10^{8}$ & $158$ &  $70$ & 9.39 \\
S3 & $80.03$ & $89.42$ & $83.85$ & $ 9.2\times10^{8}$ & $1.8\times10^{8}$ & $116$ &  $49$ & 9.71 \\

\hline

\end{tabular}

\label{tab:resims}
\end{table*}

In order to compare the physical result obtained in void resimulations, we also run a periodic box with $125$ \mph of side length (see table \ref{tab:resims}), including the same astrophysical subgrid models and the same resolution as the void resimulations, and use it as a reference sample of a homogeneous Universe. The use of the same resolution is required because the astrophysical subgrid models are very dependent on the resolution level, this has already been pointed out by \citet{Springel2003}. 
It is important to note that, by definition, this cosmological simulation does not include the large scale modes present on the void resimulations, it can be taken as a reference of the mean behaviour of the general Universe at the scales considered in this work. This simulation will be used to assess possible dependencies in the baryon process included in resimulated void regions.

\begin{figure*}
	\includegraphics[width=16cm]{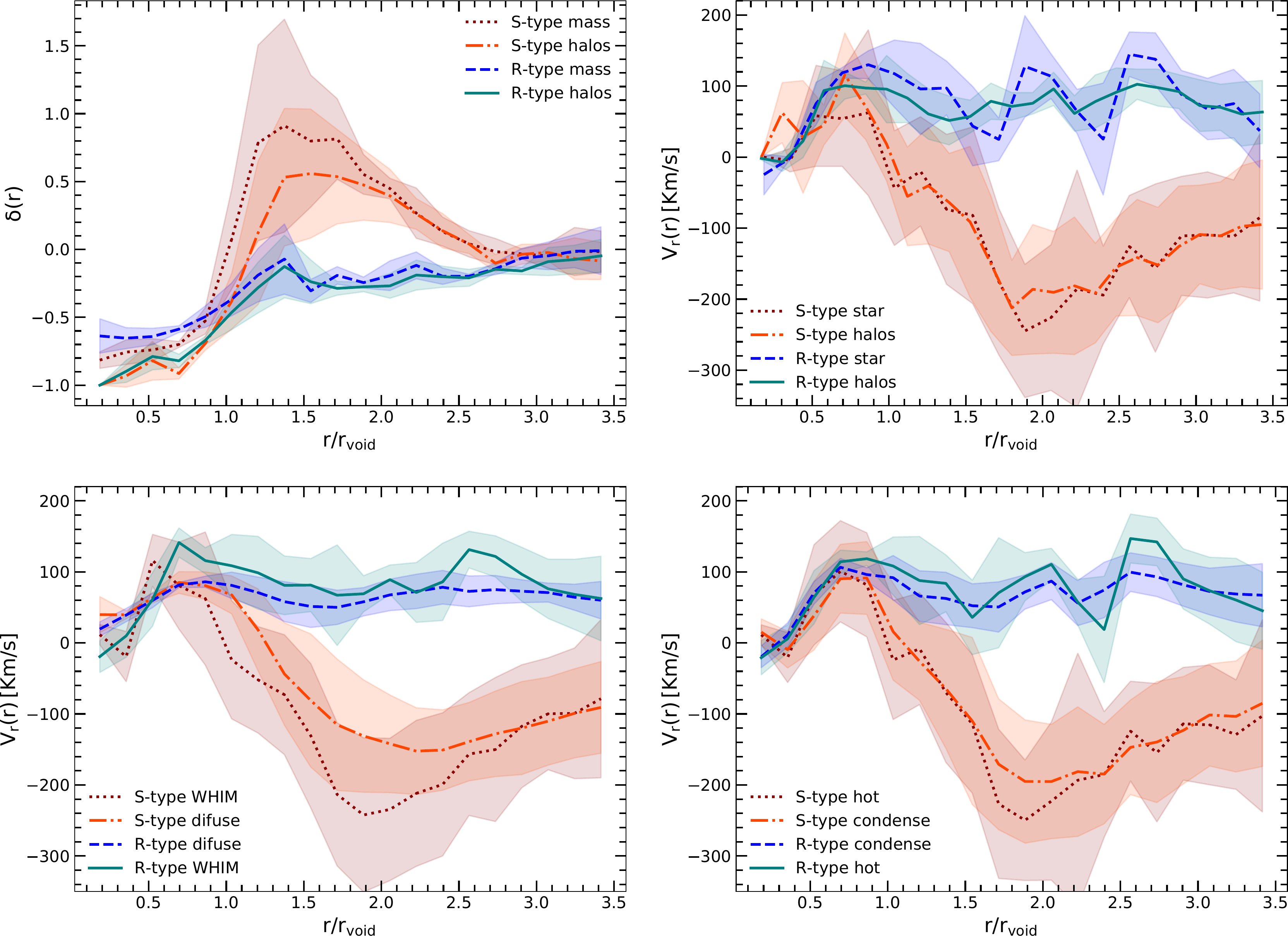}
    \caption{Stacking density and velocity profiles of voids. The different lines indicate the mean values of the stacked profile, while the shaded area shows the estimated error. 
    {\it Top left:} 
    With red doted lines we show the density contrast profile ($\delta(r)$) of the S-type voids traced by dark matter particles, while with orange dot-dashed lines the results obtained using DM haloes (with masses $M_{\rm DM}>9.2\times10^{10}$\msolh). For R-type voids, the density profiles traced by particles (haloes) are shown with blue dashed (green solid) lines.
    {\it Top right:}
    The red dotted  (orange dot-dashed) line is the mean radial velocity profile of stars (haloes) around S-type voids. Similarly, for R-type voids, we show the mean radial velocity profile traced by stellar particles and haloes in green solid and blue dashed lines, respectively.
    {\it Bottom left: } The the mean radial velocity profile traced by low density gas particles as diffuse and WHIM gas particles. The former, diffuse gas, is displayed in orange (blue) dot-dashed (dashed) line for S-type (R-type) voids. The results for WHIM gas are shown with green solid lines in green and dotted red lines for R-type and S-type voids, respectively. 
    {\it Bottom right: } Radial velocity profiles of high-density gas. Condensate gas is shown with orange dot-dashed lines and blue dashed lines, for S-type and R-type voids, respectively. Hot gas is shown using red dotted and green solid lines for S-type and R-type voids, respectively.
    }
    \label{PerfilesVoids_Resi}
\end{figure*}

\subsection{Halo catalogue and Merger Trees}
\label{subsec:haloes}

The zoom-in and the reference simulations were post-processed with the {\sc rockstar} \citep{Rockstar} halo finder, and merger trees were constructed using the {\sc ConsistentTree} code \citep{ConsistentTree}. In this work we only use the host haloes, that is, haloes not contained within any other halo and identified with at least $100$ dark matter particles within the virial radius (\rvir). Using this subsample of haloes we have verified that within the virial radius we always have at least $100$ particles of gas and therefore the hydrodynamics is well resolved by the SPH technique. 
We have also verified that the mean baryon fraction of haloes above this mass threshold is similar to the cosmological fraction ($\sim 0.18$).
We will discuss the baryon properties of haloes in more detail in section \ref{subsec:Bar_hist}.

\section{Cosmic Voids as proxy for galaxy environment properties}
\label{sec:cosm_prop}

\begin{figure*}
    \centering
    \includegraphics[width=16cm]{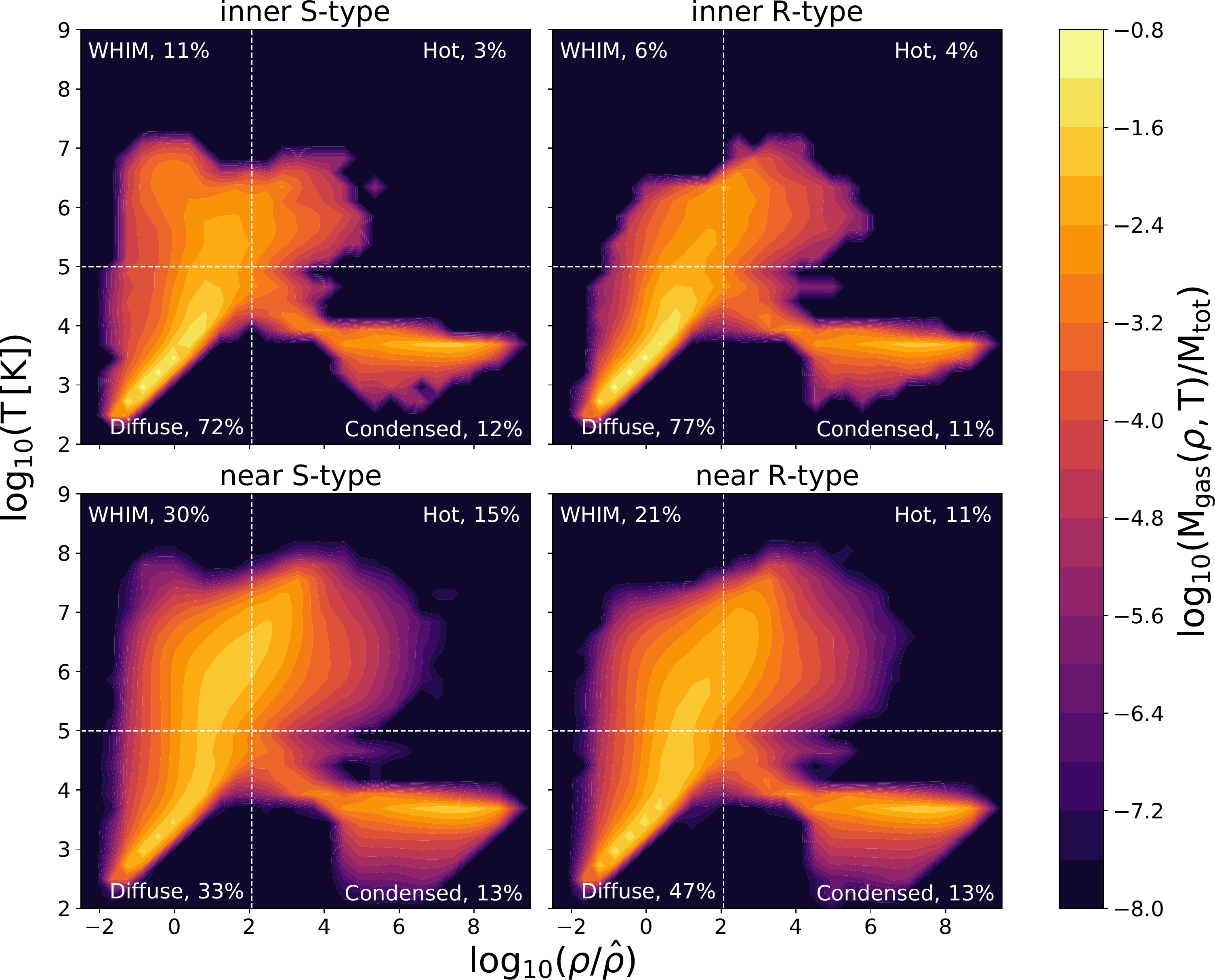}
    \caption{Stacking of density-temperature phase diagram of gas particles inside (upper panels) or near (lower panels) to voids in zoom-in simulations at z=0. Voids are stacked according to their dynamical classification, R-type and S-type corresponding to right and left panels respectively (as is indicated in figure key).  White lines indicate the threshold used to divide the gas into four phases, following \citet{Huang2019}. The colour map indicates the fraction of mass in each bin.
    }
    \label{DF_voids}
\end{figure*}

\begin{figure*}
    \includegraphics[width=17cm]{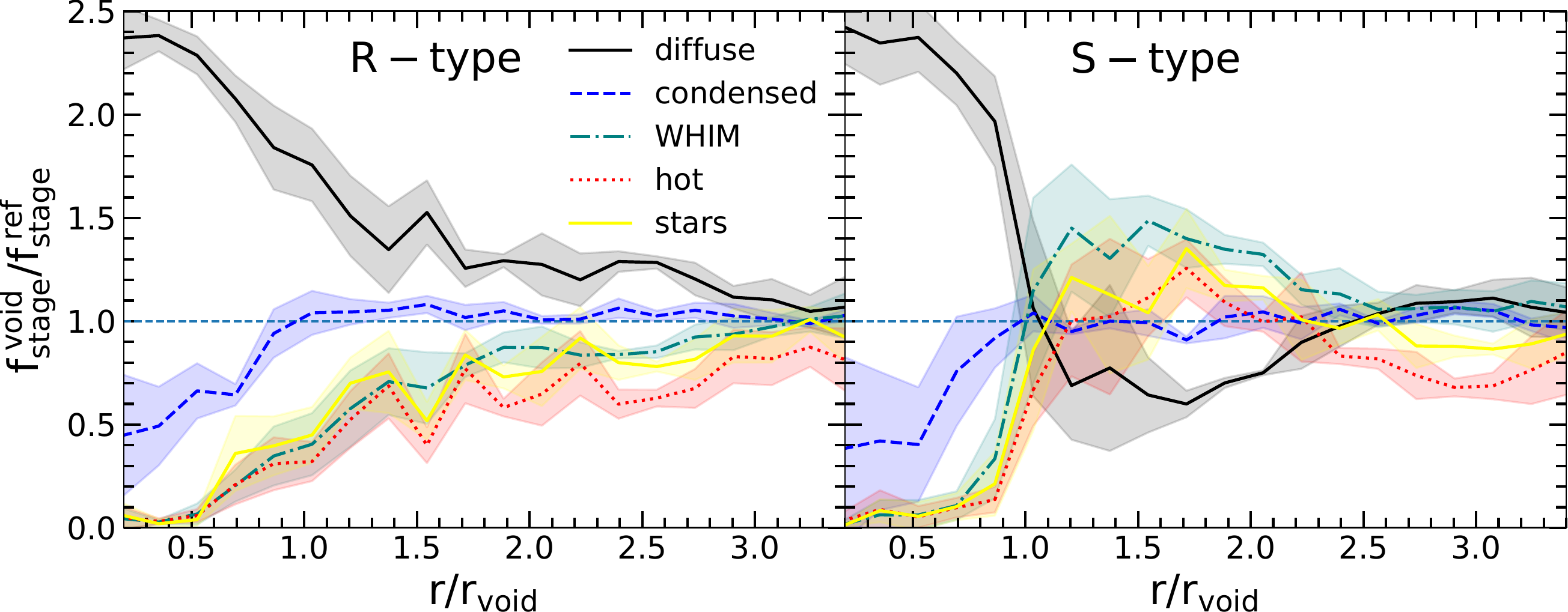}
    \caption{     Average void stacking profile of the quotient of baryon fractions at a given distance from the void centre and in the general environment of the universe (see section \ref{sec:cosm_prop}.{\it Left panel:} profile for R-type voids. {\it Right panel:} profile for S-type voids. Diffuse gas in black, condensed in blue, WHIM in green, hot in red and stars in yellow. The shaded area corresponds to the error of the mean.
    }
    \label{FraccionesRelativas}
\end{figure*}

During the large scale formation process, the baryonic matter experience a series of transitions from different thermodynamic states. These transitions define where and when stars form, depending on both the resolved astrophysics and the adopted subgrid model. 
These different states of the gas can be studied in the baryon phase diagram, which is the distribution of the gas particles as a function of their pressure and temperature. In Fig.~\ref{DF_ref} we present the reference simulation phase diagram. We divided the gas into four phases following the criterion adopted in \citet{Huang2019}:

\begin{enumerate}
    \item \textit{Diffuse} is primordial gas at low temperature and low density. It is mainly a balance between adiabatic cooling and heating by photoionization. This behaviour is characterised by an equation of state which follows a power-law relation with the density and can be seen in Fig.~\ref{DF_ref} as a straight feature in the bottom-left part of the diagram.

    \item \textit{WHIM (warm-hot intergalactic medium)} consists of gas that is shocked and heated by falling into dark matter haloes or by the star formation process, populating the large scale structures outskirts.

    \item \textit{Hot} gas which, after falling into the potential wells of the haloes, reaches high densities. In this region, we also find gas that is warmed by the thermal feedback mechanism related to the star formation process. 

    \item \textit{Condensed} gas in the inner regions of dark matter haloes that is cooled by radiative cooling mechanisms. In this region, the gas is processed so that the star formation mechanisms are activated.  
\end{enumerate}

As usual \citep{Dave2010,Martizzi2019}, we separate the phases using a limiting temperature of $10^{5}K$ and a threshold in density giving by
\begin{equation}
    \delta_{\rm th}=6\pi^{2}(1+0.4093(\Omega_{f}^{-1}-1)^{0.9052}-1)\,,\nonumber
\end{equation}
\begin{equation}
    \Omega_{f}=\frac{ \Omega_{\rm m}(1+z)^{3}}
    {\Omega_{\rm m}(1+z)^{3}+(1-\Omega_{\rm m}-\Omega_{\Lambda})(1+z)^{2}+\Omega_{\Lambda}}\,,\nonumber
\end{equation}
where  $\delta_{\rm th}$ is the virialization overdensity of dark matter haloes \citep{Kitayama1996} and $\Omega_{f}$ is the density parameter as a function of the redshift $z$. 

In this work we use the term {\it baryon stage} to refer to the baryonic matter in the form of both: gas in some of the phases defined above and stars. Each baryon stage fraction is defined as $f_{\rm stage}=\rm M_{\rm stage}/M_{\rm baryon}$, where $\rm M_{\rm stage}$ is the mass in a given stage and $\rm M_{\rm baryon}$ is the total mass of baryons, including stars and the four gas phases. The panels in Fig.~\ref{DF_ref} also show the percentage of the gas in each baryon stage (i.e. diffuse, condensed, WHIM and hot gas phases), that is $100\times f_{\rm stage}$. Due to stars are not displayed in the figure, the sum of all the percentages does not add up to $100\%$. 
Hereafter we will refer indistinctly to a given fraction or percentage of a baryon stage. We will also use superscripts to denote in which region or environment the fractions are compute: i.e. $f_{\rm stage}^{\rm void}$ and $f_{\rm stage}^{\rm ref}$ correspond to baryon fractions in voids or reference simulations, respectively. 
The $f_{\rm stage}^{\rm ref}$ values are taken as the expected results in a general universe, that is without any environmental bias and with the universal percentage of each baryon phase, restricted to physics models used here. In this way, we will study the relative fractions of baryons at different stages in voids to the general universe.

Fig. \ref{slides} shows two slices through the centre of two voids re-simulations of R- and S- types, showing the correlation between the gas properties and their location in different cosmological environments. The slice depth is one void diameter in each case (for both this is $\sim 20$ \mph). 
The interior circles indicate the void radius ($r_{\rm void}$), referred in this work as the inner void region.
The zone between this radius and the outer circle at $3\,r_{\rm void}$ is what we called the near void region. 
In different colours, we mark each of the gas phases and in black the diffuse gas. This gas phase is very abundant in the inner regions of each void and in the low-density regions in general. The density peaks, as nodes or clusters (particularly present in the S-type void shells), contain abundant hot gas (red points) produced by feedback related to the star formation processes (yellow points). 
In the surroundings of the S-type voids we find high overdensity of large scale structures, which are not so prominent in the near region of R-type voids.
The condensed phase (blue) is present in haloes while the gas cools down. In this plot is most visible in low-mass haloes in the inner and surrounding regions of voids where the star formation process is less present. 
The WHIM gas (cyan points) is typically in the environment of big filaments and structures, as stated in \citet{Martizzi2019}.

Fig.~\ref{slides} allow us to understand the characterisation between S- and R-type voids. The former has an integrated density contrast profile that is overcompensated due to large structures in its walls \citep{Ceccarelli2013, Paz2013}. R-type voids are not overcompensated at any scale and its integrated density contrast profile is growing steadily until reaches the mean density of the Universe. 
Fig.~\ref{PerfilesVoids_Resi} shows in the top-left panel the differential contrast density profile for haloes with $M_{\rm DM}>10^{11}$~\msolh and for the dark matter (labeled as mass). 
The differential density contrast is defined as
\begin{equation}
    \delta(r) = \frac{n(r) - \hat{n}}{\hat{n}}\,,
\end{equation}
where $n(r)$ is the halo or matter density in a shell with inner radius $r$ and outer radius $r+dr$ and $\hat{n}$ is the mean density of haloes or particles. The mean density is calculated with the reference simulation.
These profiles are calculated by stacking all voids of each type. The shaded regions correspond to the variance with respect to the mean of the stack. S-type voids show in the near region a contrast greater than zero which means that the differential densities are greater than the mean density of the Universe. 

\begin{figure}
	\includegraphics[width=\columnwidth]{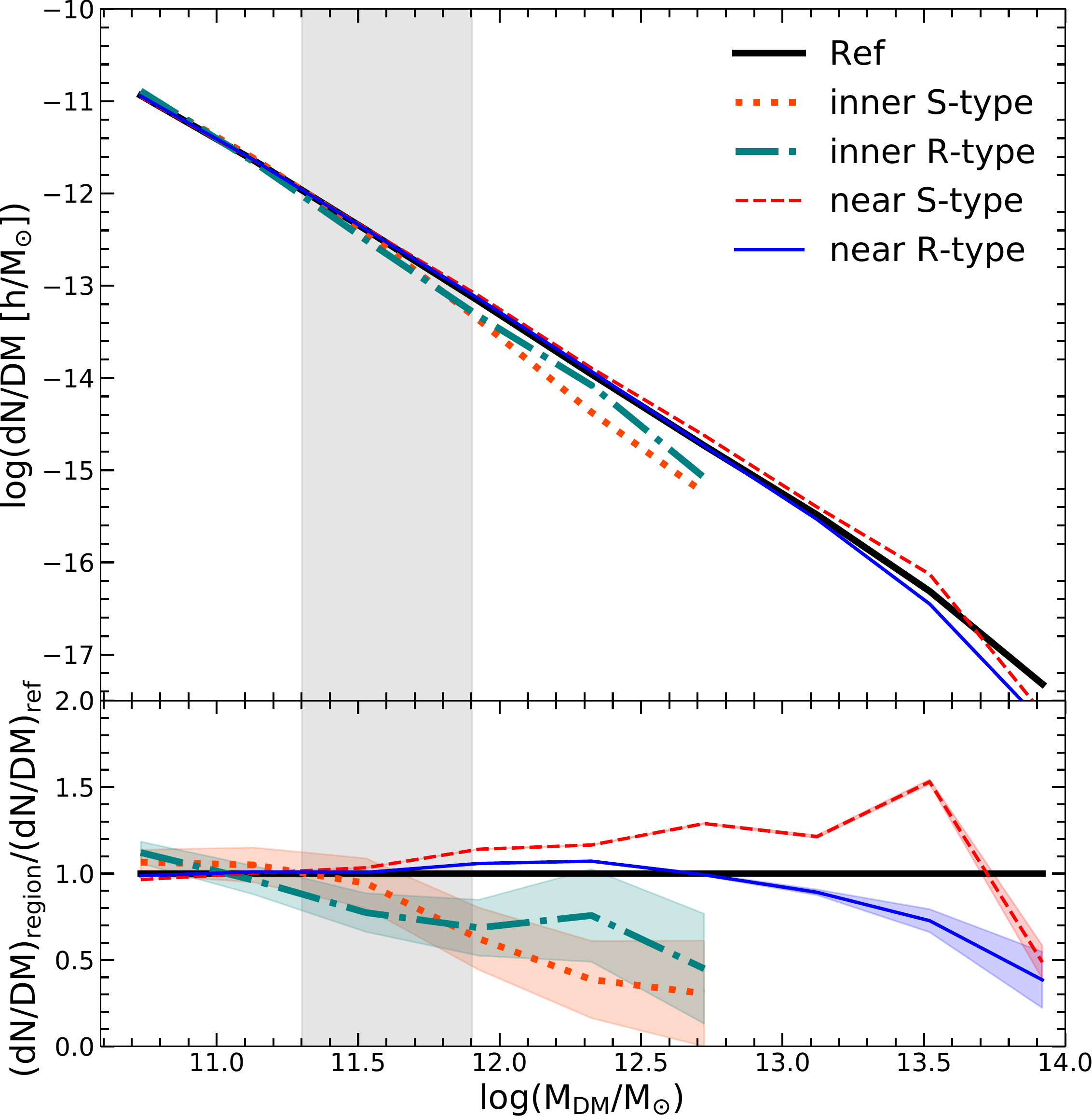}
    \caption{ 
       Halo mass function in different environments. In black solid line are the results for the reference simulation. The mass distribution of haloes inside and in the near regions of R- or S-type voids are shown with different colours and line styles, as indicated in the figure key. In orange dotted (red dashed) lines is the distribution in inner (near) regions of S-type voids. In green dot dashed (blue solid) lines the inner (near) R-type void region result is shown.  The shaded area shows the uncertainty in the mass distribution. The grey shaded region indicate the halo range in mass analysed in this work ($2\times10^{11}h^{-1}\rm M_{\odot}<M_{\rm DM}<8\times10^{11}$$h^{-1}$$\rm M_{\odot}) $}
    \label{FuncionesDeMasa}
\end{figure}

The rest of the panels (Fig.~\ref{PerfilesVoids_Resi}) show the radial mean velocity profile of haloes and stars (upper-right), diffuse and WHIM gas (lower-left) and condensed and hot gas (lower-right). In each panel, the shaded area indicates the variance between different voids in our sample. 
The mean profile is calculated by computing the radial velocity of each particle with respect to the mean motion of its corresponding void. 
The void velocity is defined as the mean velocity of the haloes between $0.8 - 1.2\, \rm r_{void}$ on the resimulations. It is important to note, that the mean velocity of haloes correlates with the particle mean velocity, as was demostrate in  \citep{Lambas2016}.
As can be seen in Fig.~\ref{PerfilesVoids_Resi}, the different characteristic behaviour of R- and S-type voids are shown by the expansion or contraction dynamics of the shells in each type \citep{Ceccarelli2013}.
Despite minor differences in the average profile of stars, haloes and gas phases, the profiles are consistent.
As a result, we can conclude that R/S-type voids, not only show a different dynamic of expansion and contraction, but also have different distributions of baryons.

To quantify this we construct the phase diagrams of gas inside one void radius and in the near void region ($1-3\,r_{\mathrm{void}}$). This is performed by taking all gas particles inside these two regions, for the different resimulated voids accordingly to the void type. The results by stacking all the voids of a given type are shown in Fig.~\ref{DF_voids}. As in Fig.~\ref{DF_ref}, we show the baryons fraction $f_{\rm bar}$ in the different environments. 
In general, phase diagrams in the inner region can be characterised by an absence of gas in the WHIM and hot phases (upper panels). In contrast, in the near region of S-type voids, there is an excess of hot gas with respect to the reference box (Fig.~\ref{DF_ref}) and the stacking of R-type voids. This is due to the large structures present in these regions with abundant stars that warm up the gas by feedback and the accretion shocks when gas falls to potential centers (Fig.~\ref{slides}). When considering the inner region in Fig.~\ref{DF_voids}, we observe that the baryons do not differ much, but for a slight excess in the WHIM material in the S-type void.

In order to characterise the baryon spatial distribution accordingly to its state, we define the relative baryon fraction for each stage as the quotient between $f^{\rm void}_{\rm stage}$ at a given radial shell and the universal fraction $f^{\rm ref}_{\rm stage}$, obtained from the reference simulation (see Fig.~\ref{DF_ref}). 
These profiles are shown in Fig.~\ref{FraccionesRelativas} by stacking voids of S/R types, the shaded areas are the corresponding variances of each void sample. 
The inner regions of the voids are composed of abundant diffuse gas and scarce hot-WHIM gas and condensate. This can be understood in terms of the low density of these environments indicating little evolution of the structure. This gas is mainly cold, at low densities, and can be considered as a reservoir for future structure formation. The Fig. \ref{FraccionesRelativas} show that the small excess of WHIM gas present in inner region of S-type voids (see Fig. \ref{DF_voids}) is located at $\sim \rm r_{void}$, where the near to S-type void region begins. The relative fraction profiles of the voids are very different depending on their dynamic behaviour.
R-type voids (left) decrease their diffuse gas fraction in order to increase their condensed and hot-warm fractions.
Close to $\sim3\,r_{\rm void}$ fractions reach the cosmological mean, indicating that the influence of voids extends far beyond the radius at which they were identified. The behaviour of the relative fraction profile of the S-type voids (right) has an abrupt change of behaviour when reaching the void radius. 
The hot-WHIM gas fractions exceed the cosmological mean and the diffuse gas is scarcer. This is understandable in terms of the over-abundance of haloes present in the near region of these voids. 
The void integrated density contrast profile is overcompensated and for this reason, large structures such as groups and clusters form in the nearby region, containing large numbers of stellar particles which are warming up by feedback the surrounding environment. 
In Fig.~\ref{PerfilesVoids_Resi} the near regions, recall they are defined between $\sim 1-3\,r_{\rm void}$, has a density above the cosmological average. This is in a good agreement with the regions with an overabundance of hot and warm gas in Fig.~\ref{FraccionesRelativas}. 
Similar to what happens in the case of R-type voids, in S-type voids it can be seen that towards $\sim 3\,r_{\rm void}$ the fractions of each gas phase tend to reach the cosmological mean. 
Notably, for both void types, the condensed gas shows the same behaviour: monotonically increases to the cosmological mean around $r_{\rm void}$, showing no evolution at larger scales. 
This is mainly due to the feedback models implemented in this work.
Condensed gas is naturally located inside haloes, in the inner void regions there are very few haloes and they have low masses. 
However in the near region for both void types, halo abundances are around or above the mean of the Universe. Therefore, a significant fraction of massive haloes is present, where the star formation process occur more efficiently.
In this way, the star formation mechanisms steadily transform diffuse gas to hot gas and stars passing through the intermediate condensed gas phase. In the near region of both void types in Fig.~\ref{FraccionesRelativas} it can be seen a balance in this process that keeps the condensed gas phase around the mean Universe fraction.
It is important to note that due to the resolution of our simulations, gas in the condensed phase is unlikely to be found in haloes with masses below $\sim10^{11}$ \msolh, as the number of particles is not sufficient to resolve the SPH correctly.

In Fig.~\ref{FuncionesDeMasa} we show the mass distribution of dark matter haloes. The coloured shaded areas represent the Poisson uncertainty in the mass distribution, whereas the grey shaded vertical strip is the mass range where the fraction of condensed gas is significant, the lower limit is imposed by the resolution (200 particles as will be used in the following analysis, see section \ref{sec:DM_haloes}) whereas the upper limit is where the condensed fraction becomes less dominant. 
In this range the mass distribution of haloes in near regions of voids and reference simulation is similar. 
This explains to some extent the similarity of the behaviour of the relative fraction in the near region of S-type and R-type voids.

The results of these sections show that the distribution of baryons depend on the environment defined by the void density profile. We show that the distance to the void centre is a good proxy for baryon properties and that the void dynamics (R/S-type) is also important.
Mainly the differences manifests themselves in the near voids region, as pointed out in Fig.s~\ref{PerfilesVoids_Resi}, \ref{DF_voids} and \ref{FraccionesRelativas}.
The differences found appear to be a manifestation of the different behaviour of the mass distribution (Fig.~\ref{FuncionesDeMasa}) since the baryon distribution in its different states reflects a greater or lesser degree of processing that occurs in the haloes.  
In the next section, we focus on the evolution of the haloes and the impact this has on the distribution of baryons. 

\section{Haloes evolution}
\label{sec:DM_haloes}

As we discussed in the previous section, the baryon distribution manifests dependencies on the environment, therefore we analyse how the void ambient affects the formation and evolution of galaxies and haloes.
We separate the halo sample accordingly to the void type (R/S) and region, that is inner (located inside the void radius $r_{\rm void}$) and near (at void centric distances between $1-3\,r_{\rm void}$). 
The top panel of Fig. \ref{FuncionesDeMasa} shows the dark matter haloes mass distribution in these regions separated by void type and the reference simulation for comparison. The mass distribution of haloes in the reference simulation is shown with a black solid line. For haloes in the inner void regions of R and S type voids we use dot-dahsed green and doted orange lines, respectively. In the case of near regions, with blue solid line are the results for R-type voids while in red dashed lines the corresponding to S-type void regions. The colour line type scheme used is also indicated in the figure key.
The mass distributions show significant differences between the halo sample in near regions, of both void types, and the reference simulation halo sample, in particular in the high mass end. 
In general, the inner regions to voids show a distribution that falls at masses higher than $\sim10^{12}$\msolh. This is an expected behaviour due to voids are identified as underdense regions (below $10\%$ of the mean density) in a sample of haloes with masses greater than $1.4\times10^{12}$\msolh.
The near regions show differences with the void type. 
In near regions of S-type voids, the mass distribution is larger at high masses in comparison to the reference simulation. This is consistent with the presence of big structures in this type of voids \citep{Lares2017}.
On the other hand, the near region of R-type voids is underdense respect to the mean density of the universe, therefore it is expected that the mass distribution of haloes falls at high masses. This can be seen in Fig. \ref{FuncionesDeMasa} when comparing the R-type voids and reference mass distributions.
The grey shaded area indicate the mass interval selected for the analyses presented in the following sections. The lower limit is imposed by a resolution criteria: we consider only haloes with more than $200$ dark matter particles (for more details see in sub-section \ref{subsec:Bar_hist}). 
The upper limit corresponds to $8\,\times 10^{11}$ \msolh, and is imposed in order to have comparable masses along all halo subsamples, that is haloes in different void types and regions. This allows us to analyse possible dependencies of galaxy and halo properties independent of halo mass. 
As consequence we obtain a total sample of haloes in the near region of R (S) type voids of $3309$ ($3593$) whereas for the inner region $58$ ($54$) haloes. Even though these numbers are similar, given the different number of void resimulations of each void type, the mean number of haloes in both regions is larger for S-type voids. In the reference simulation at this mass interval, we have $22559$ haloes. 

\subsection{Dynamical states of dark matter haloes}
\label{subsec:dyn_states}

Motivated by the results shown in the previous section, here we present a study of the inner properties of haloes in different void regions. 
We compute the kinematic phase-space diagram of haloes to characterise their dynamics and evolution accordingly to the previously defined void regions and comparing them with the reference simulation. 
As we mention at the end of the previous section we restrict our analysis to haloes with masses between $2\times10^{11}$\msolh$<\rm M_{DM}<8\times10^{11}$ \msolh. 
Phase-space diagrams will be computed for different samples of haloes in this interval, separated accordingly to its membership to the inner or near regions of a given void type.
The phase-space distribution of particles around haloes is defined in bins of the radial component of the particle velocity in the rest frame of the central halo ($v_{\rm r}$) and its distance to the centre of mass of the halo ($\rm r$). The distances are defined in units of the halo virial radius, that is $\rho = r/r_{\rm vir}$, while the velocities ($\nu$) in units of halo circular velocity, i.e. $\nu = v_{\rm r}/v_{\rm circ}$ (where $v_{\rm circ}=\sqrt{GM_{\rm vir}/r_{\rm vir}}$).
Finally, we compute the mean phase-space diagram of a given halo sample as the sum of all halo particle counts dividing by the number of haloes, hereafter denoted as $N(\rho,\nu)$.

\begin{figure*}
    \centering
    \includegraphics[width=\linewidth]{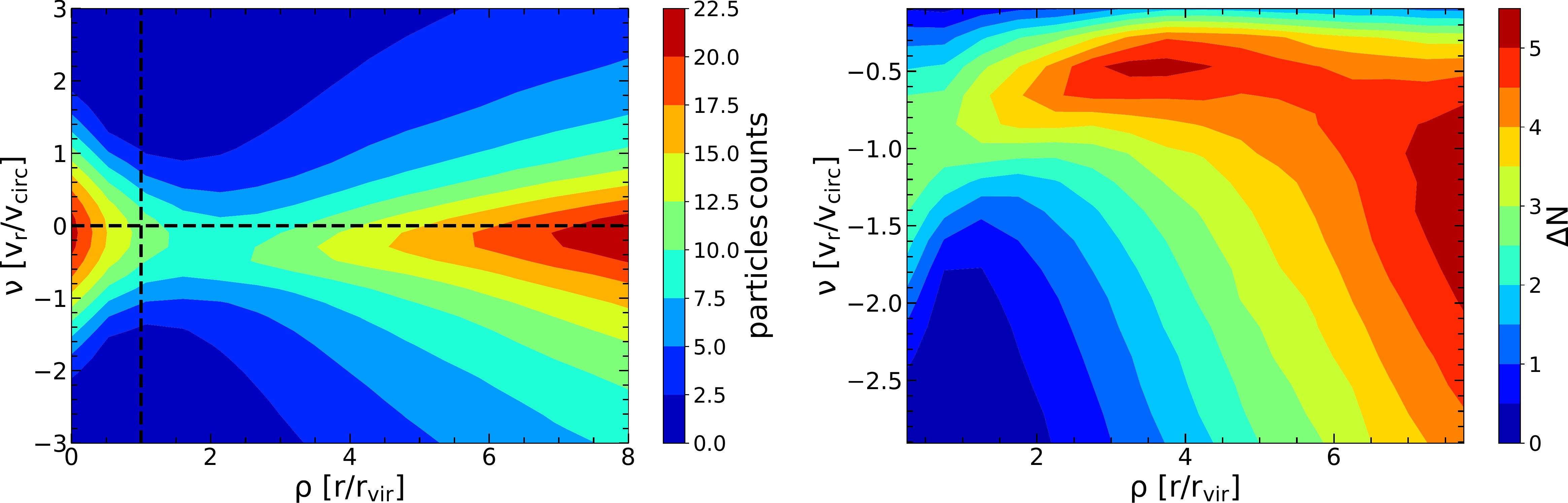}
    
    \caption{Left panel: kinematic phase-space diagram for dark matter particles around haloes with masses between $2\times10^{11}$ \msolh to $8\times10^{11}$\msolh in the reference simulation. The vertical line delimit the virial region. The horizontal line divide the diagram into particles with positive velocities and negative velocities. The right panel shows the mirrored phase-space $\Delta N$ map (as defined in section \ref{sec:DM_haloes}).     }
    \label{EspacioFasesInicial}
\end{figure*}

As an example of this kind of diagrams, we show in the left panel of Fig.~\ref{EspacioFasesInicial} the corresponding kinetic phase-space diagram of haloes from the reference simulation. This map can be taken as a reference of the behaviour of haloes in the mass range of interest regardless of the halo location with respect to void centres. 
Similar kinetic phase-space diagram has been well studied in previous works \citep[][ and references therein]{Busha2005,Cuesta2008}. In this work we use it to compare the results in the reference simulation with the corresponding maps for haloes in voids. 
 The region inside the virial radius is comprised of particles having zero average radial velocity. This virial region can be clearly seen in Fig.~\ref{EspacioFasesInicial} as an overdensity of particles inside $\rho=1$ in the left part of the diagram, distributed symmetrically in radial velocities. On the other hand, the infall region consists of particles beyond \rvir  ($\rho>1$) that are in process of being accreted by the halo. This region can be seen as an asymmetric structure of particles respecting the horizontal axis at $\nu=0$, where there is an excess tail of particles with negative velocities (therefore infalling).

To better show the differences between the infall and outfall of mass particles, we look for the asymmetries in the distribution around the null radial velocity, indicated by the horizontal line in the Left panel in Fig~\ref{EspacioFasesInicial}. To do this we compute the map $\Delta N$ as the difference between the number of infall and outfall particles at a given distance $r$ and at a given bin of absolute radial velocity, i.e. $|\nu|$. In other words, this maps is simply defined as $\Delta N(\rho,|\nu|) = N(\rho,-|\nu|)-N(\rho,|\nu|)$, which can be interpreted as vertically mirroring the upper part of the phase diagram (corresponding to outfall particles) and then subtract it to the lower part of the diagram (corresponding to infall particles).

This is shown in the right panel of Fig.~\ref{EspacioFasesInicial}. As it can be seen the values of $\Delta N$ are predominately positive indicating that in general particles around haloes are in infall. To emphasise this fact, we decided to express the velocity axis with negative values. This map reveals in a more clearly way the structure of the infall of matter. Moreover, the symmetric structure of particles corresponding to the virial region has almost disappeared in this panel, indicating that in the virialized region infall and outfall particles are in equilibrium. For $\rho>1$, it can be clearly distinguished two peaks in the $\Delta N$ map, revealing that there are two components in the net infall of matter. The component in the region $1<\rho<6$ corresponds to particles at intermediate distances infalling to the haloes with moderate velocities. This component can be thought of as particles coming from the neighbouring filaments. Therefore we decided to call this component the proper net infall of matter. Whereas the component at larger distances, $\rho>6$, has a broad distribution of high velocities, corresponding to the infall of the halo frame into larger structures (e.g. clusters of galaxies). We call this second component simply as the large scale flow. 

\begin{figure}
    \centering
    \includegraphics[width=\linewidth]{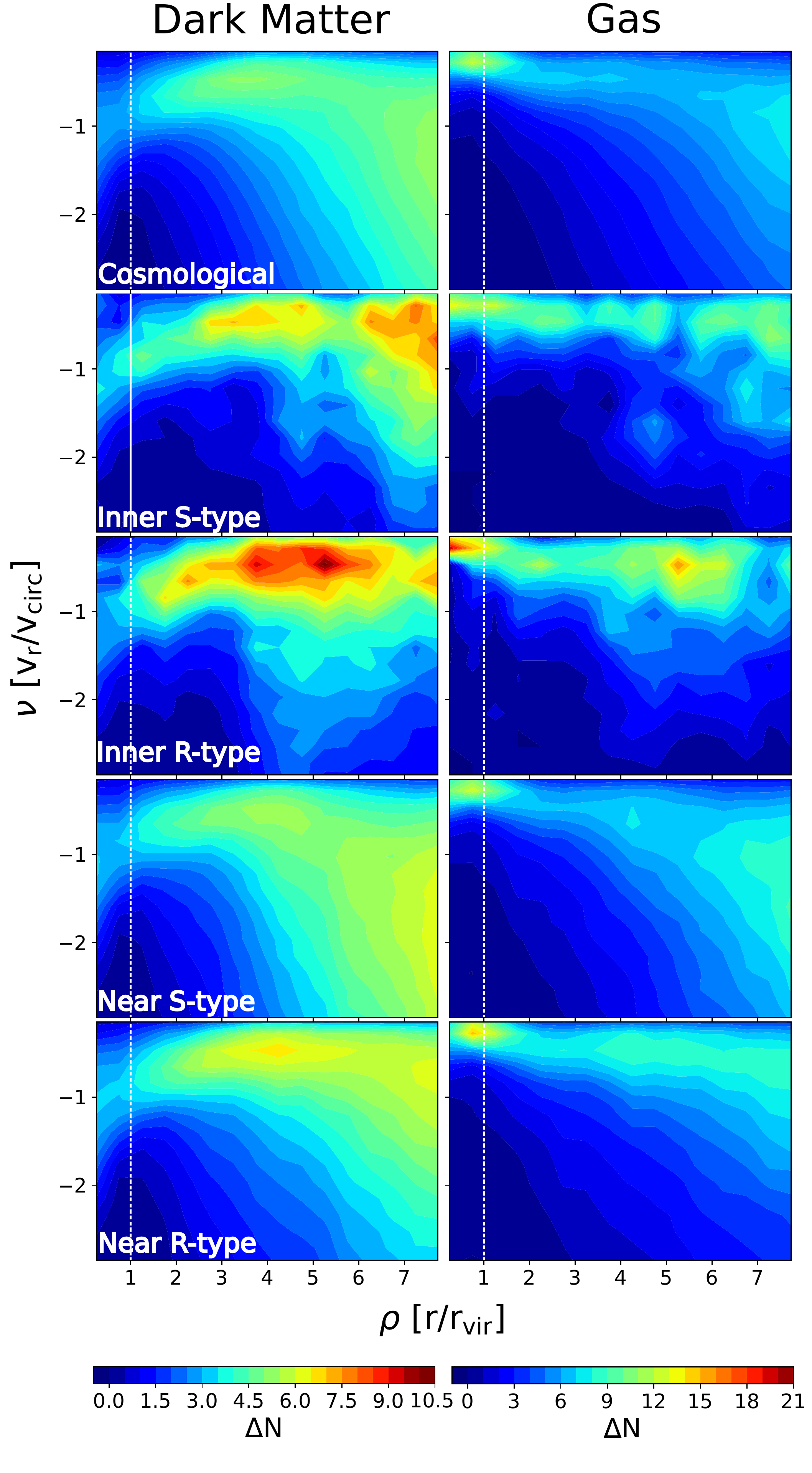}
    \caption{ $\Delta N$ maps (see section \ref{sec:DM_haloes}) showing the structure of the infall of matter around haloes.
    The set of panels arranged in the left column show maps in dark matter particles around haloes in different environments, while in the right column are the equivalent maps but computed using gas particles. 
    From top to bottom are results for different environments: in the top the reference simulation, below the inner regions of S-type and R-type voids and in the last two bottom panels are near regions of S-type and R-type voids. Different colours indicate the variation in the number excess of infalling particles $\Delta N$. 
    haloes have virial masses between $2\times10^{11}$ \msolh to $8\times10^{11}$\msolh
    }
    \label{EspacioFases}
\end{figure}

As we mention at the beginning of this section, we are interested in the study whether it could be dependence in the halo dynamics with the void type and environment.
With this aim we have constructed the $\Delta N$ maps for different subsamples of haloes selected accordingly to the void type and whether they belong to the near or inner regions. 
This is shown in Fig.~\ref{EspacioFases}, where the set of plots in the left column corresponds to the $\Delta N$ map of dark matter particles around haloes, whereas the plots arranged in the right column corresponds to the results obtained using gas particles. From top to bottom we have $\Delta N$ maps for haloes: in the reference simulation, in the inner region of S-type, in the inner region of R-type and in the near regions of S-type and R-type voids, as it is labeled in the figure. 
The colour-map shows that the component of proper infall is more pronounced for haloes inside of voids. This is particularly stronger in haloes living in R-type voids. The large scale flow seems to be larger for haloes inside of S-type voids in comparison with haloes inside of R-type voids. The behaviour in the near void regions is also dependent on the void type.  Analogously to the case of inner regions, in the near case we have a more pronounced proper net infall in R-type voids. 
This is because the near region of R-type voids is also underdense respecting to the general universe. Then this infall component could be associated with the fact that this haloes inhabits an underdense environment. 
The near region of S-type voids is highly over-dense and contains massive structures which highly affect the halo velocities. This effect is shown as an excess of $\Delta N$ in the large scale flow component. As it can be seen from the mass distribution (Fig.~\ref{FuncionesDeMasa}), there are lesser massive haloes in the near region of R-type voids compared to the reference simulation. This reflects in a smaller infall associated with the large scale flow component.
Also in comparison with the reference simulation, near regions of S-type voids have an excess of massive structures, producing an excess of infall at large scales. 

As we mentioned in the previous paragraph, in the right column of Fig.~\ref{EspacioFases} we show $\Delta N$ maps of net infall of gas particles around haloes in different environments. These were calculated in the same way as for dark matter particles. 
The general trends are similar, we have a more pronounced infall of matter inside voids and the same difference between near regions of R-type and S-type voids. 
This indicates that the accretion of gas is dominated by the dark matter behaviour. It is important to stress the fact that the $\Delta N$ map shows the difference in the number of particles, therefore in order to do a quantitative comparison between both sides of Fig.~\ref{EspacioFases} it would be necessary to weight each count by the particle mass. However, in our analysis, we are interested in the behaviour of matter infall depending on the environment and the qualitative differences between gas and dark matter. As general behaviour of the gas, we observe that in the infall region the velocity is decreasing towards the halo centre. This could be related to the effects of viscosity and pressure. In the virialized region the cooling process becomes effective and the density of gas particles increases, with lower radial velocity component. 
Coming back to the case of dark matter particles (left column panels of Fig.~\ref{EspacioFases}), we see an excess of $\Delta N$ at higher infalling velocities inside the virial radius. This asymmetry corresponds to a small fraction of particles infalling more recently into the halo, and therefore not reaching yet the equilibrium established by dynamical friction in the virialized region. 
Regarding the large scale flow component we observe similar behaviour in both dark matter and gas diagrams.

\begin{figure}
    \centering
    \includegraphics[width=\linewidth]{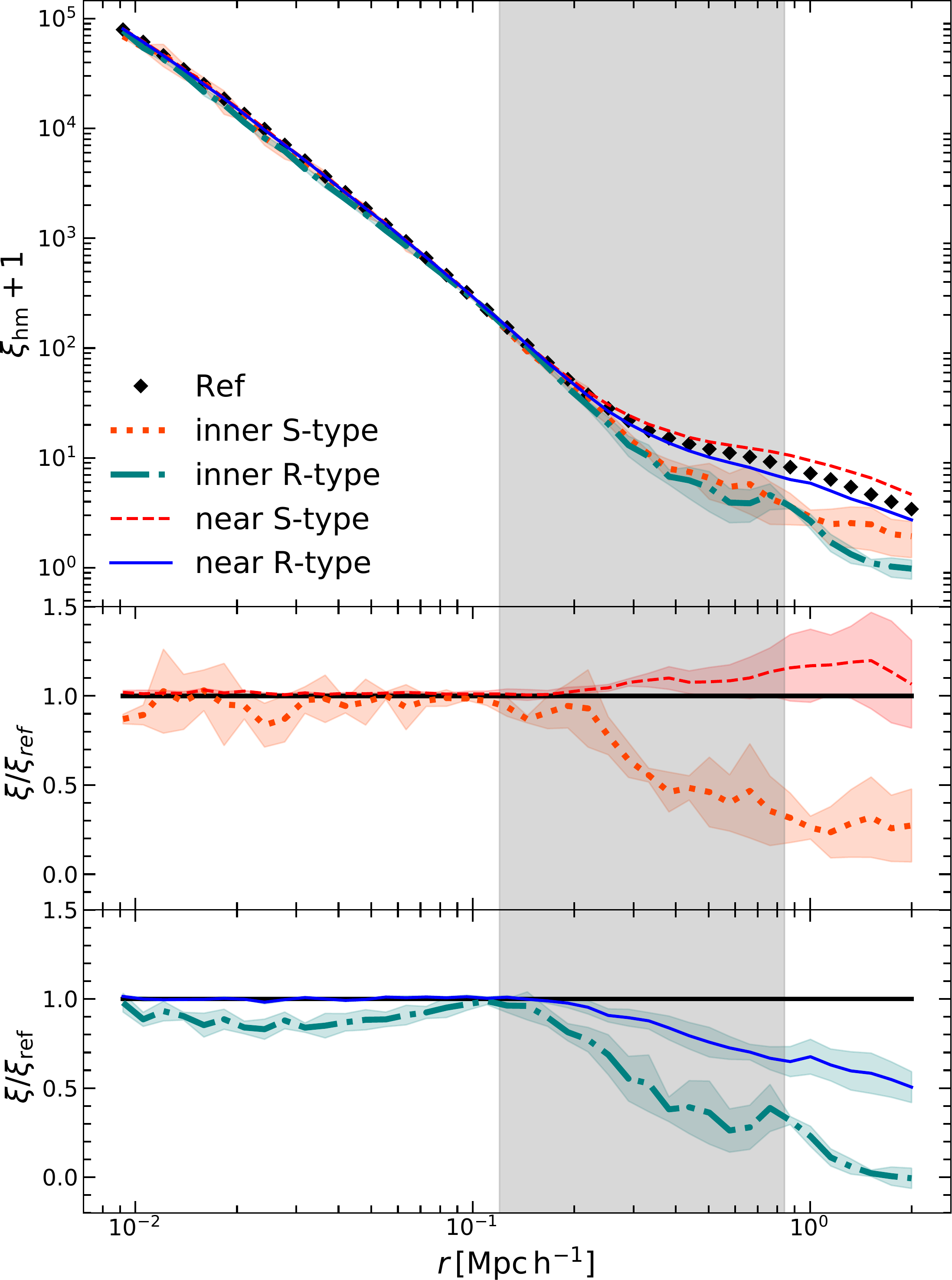}
    \caption{ Top panel: halo-particle correlation functions, $\xi_{\rm hm}$, for different environments. In black diamonds is the halo particle correlation in the reference simulation. The blue solid and red dashed lines are the correlation for haloes in the near region of R-type and S-type voids, respectively. The teal dot-dashed and orange dotted lines show the results in the inner region of R-type and S-type voids, respectively. 
    Central and bottom panels: the quotient of correlation functions in different environments and the reference simulation. In the central panel are the results for S-type void regions and in the bottom panel for R-type void regions. The halo mass range is $2\times10^{11}$ \msolh to $8\times10^{11}$\msolh
    }
    \label{correlaciones}
\end{figure}

In the top panel of Fig.~\ref{correlaciones} we show the halo-particle correlation function, where halo masses are in the range $2\times10^{11}$ \msolh to $8\times10^{11}$\msolh, as mention before. These correlation functions are estimated by counting halo-particle pairs at different bin of distances, normalising by the bin volume, the number of haloes and the particle density. The shaded areas are the variance for each correlation measure over different void resimulations. The middle (bottom) panel shows the halo-particle correlation functions for haloes in the inner and near region of S-type (R-type) voids divided by the correlation in the reference simulation.
As can be seen, up to $0.2$ \mph the correlation functions corresponding to the near region of R and S-type voids are very similar to the cosmological reference. In other words, the clustering in the one-halo term seems to be independent of the void type in near regions. However, in the case of haloes in the inner region of voids, there is a tendency to have smaller amplitudes of clustering in the one-halo term, more notably in the R-type inner region.
At large scales, that is in the two halo regime, there is a lower amplitude in the correlation function of haloes inside both types of voids, indicating that these haloes are less clustered with the surrounding structures. 
In the case of haloes in near regions, there is a different behaviour according to the void type. As expected, haloes in the near region of S-type voids are much more clustered than in the reference simulation whereas the opposite behaviour is seen for haloes near R-type voids. The vertical lines indicate the scales associated with $r_{\rm vir}$ and $6\,r_{\rm vir}$. These lines are placed to handle the comparison with the structure of infall in phase diagrams, as can be seen in Fig.~\ref{EspacioFases}. The proper infall components (the $\Delta N$ peak between $1-6\,r_{\rm vir}$) increase as the clustering in this region decreases. In other words, minimum values of proper infall component are seen in the case of haloes in the near region of S-type, while their clustering is larger than the reference. The opposite behaviour is seen in the inner region of R-type voids, haloes there have a clustering at these scales smaller than in the reference. 
Regarding the large scale flow component, that is at distances larger than $6\,r_{\rm vir}$, the relation between the amplitude of the $\Delta N$ map and the clustering magnitude behaves differently. The sample with larger clustering, that is haloes in the near region of S-type voids, have the largest component of large scale flow infall, while the opposite is seen for haloes in the inner R-type. 
These different correlations between the clustering and the components of matter infall can be thought in terms of the different nature of both components. 
As was mentioned before, the large scale flow component is related with the infall of the halo frame into higher mass surrounding structures, at larger scales. 
On the other hand, the proper infall component is related with the infall into the halo potential well of the surrounding matter.
Lower clustering at a given scale implies that the halo sample is likely to be representative of the largest structures at those scales, having then a larger $\Delta N$ component in their surrounding while these haloes have typically lower velocities in the large scale structure frame. Consequently, these haloes also have a smaller large scale flow component. The opposite is expected for haloes with large clustering, they will be more likely infalling into larger structures, and therefore its influence in their surroundings ($1-6\, r_{\rm vir}$) is suppressed.

\subsection{Baryon stages evolution history inside haloes}
\label{subsec:Bar_hist}

\begin{figure*}
	\includegraphics[width=\linewidth]{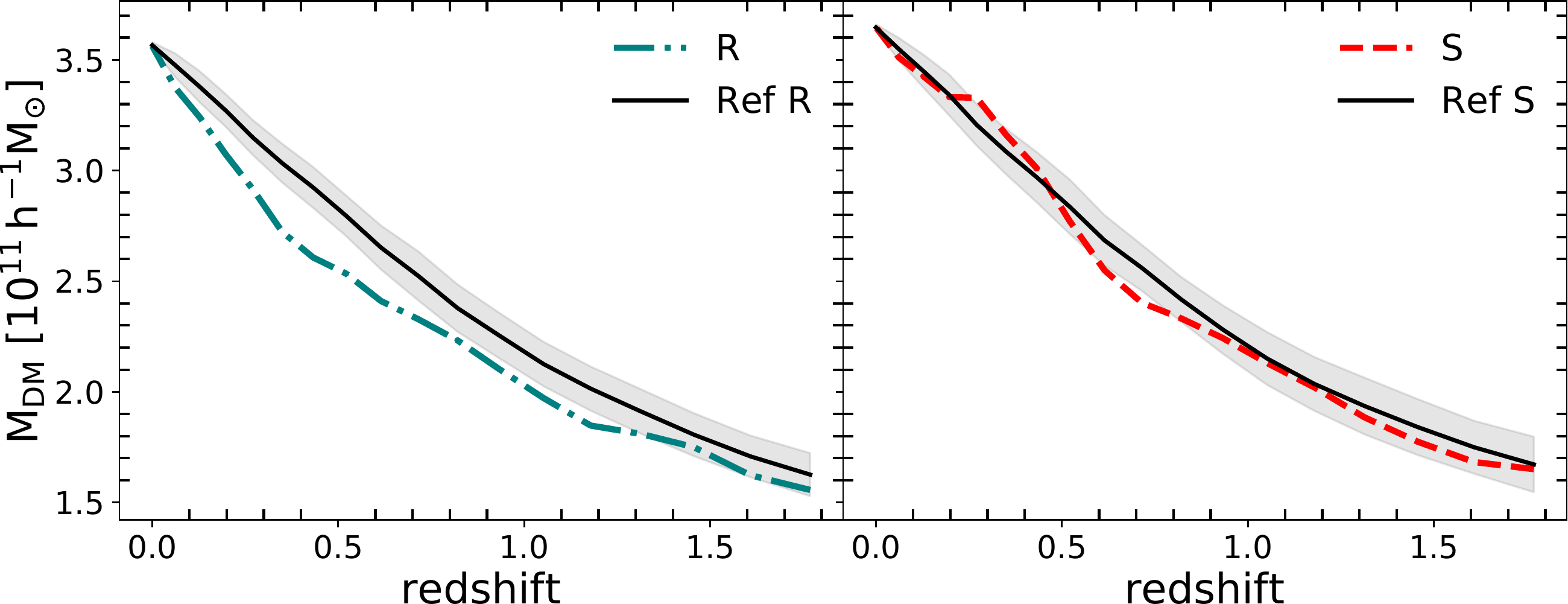}
    \caption{ Evolution of the mean virial mass of haloes in the mass range of $2\times10^{11}h^{-1}M_{\odot}<\rm M_{vir}<8\times10^{11}h^{-1}M_{\odot}$. In the left and right panels are the results for haloes in R-type voids (blue dot-dashed) and haloes in S-type voids (red dashed), respectively. Black solid lines are the results for halo samples selected in the reference simulation with an equivalent mass function to the halo sample in each void type. The shaded area is the normal deviation of the mean.}
    \label{trees}
\end{figure*}

In this section we study the evolution of the different baryon stages in haloes as selected in previous sections, that is matter in the form of stars and gas in four different phases: Diffuse, WHIM, Hot and Condensed (see section \ref{sec:cosm_prop}). Once again we restrict our analysis to haloes with masses between $2\times10^{11}$ \msolh to $8\times10^{11}$\msol. By computing the merger tree of each halo we follow the main branch up to redshift $2$. This high redshift limit is imposed in order to have enough resolution inside haloes, that is at least $100$ particles (see section \ref{subsec:haloes}). 

In order to make a fair comparison between haloes in the reference simulation and those in void resimulations we construct comparable halo mass distribution at $z=0$. 
We achieve this by selecting randomly a sample of haloes in the reference simulation in order to mimic the mass distribution of haloes at $z=0$ in the void resimulations. 
Due to a large number of haloes in the reference simulation, we can produce several samples with the same number of haloes in void resimulations with the same mass distribution. In particular, we generate $100$ realisations for each void type.
Therefore we are able to compute the standard deviation ($\sigma$) in the dark matter mass evolution and use it as an estimate of the expected error of the sample of haloes in void, $\epsilon=\sigma/\sqrt{n}$.
Fig.~\ref{trees} shows the dark matter mass redshift evolution for haloes in voids and those selected in the reference simulation. 
On the left (right) panel we have the results for haloes in R(S)-type voids, drawn in blue dot-dashed (red dashed) lines. 
In both panels, the black solid line represents the corresponding evolution in the reference simulation, where haloes have been selected accordingly to its mass as described above. The error in the mean, $\epsilon$, is shown as the shaded grey area.
As it can be seen, the evolution of halo mass on the main branch is sensitive to the void type and region. For instance, haloes in the inner region of R-type voids evolve more slowly in comparison to those in the reference simulation. In this direction, R-type haloes are less massive in comparison with the Universe on all the redshift range analysed. This difference can be in the order of $\sim 3\,\epsilon$ at $z\sim0.5$. Haloes inside of S-type void show similar trends to the R-type void haloes, but with small significance.

\begin{figure*}
	\includegraphics[width=\linewidth]{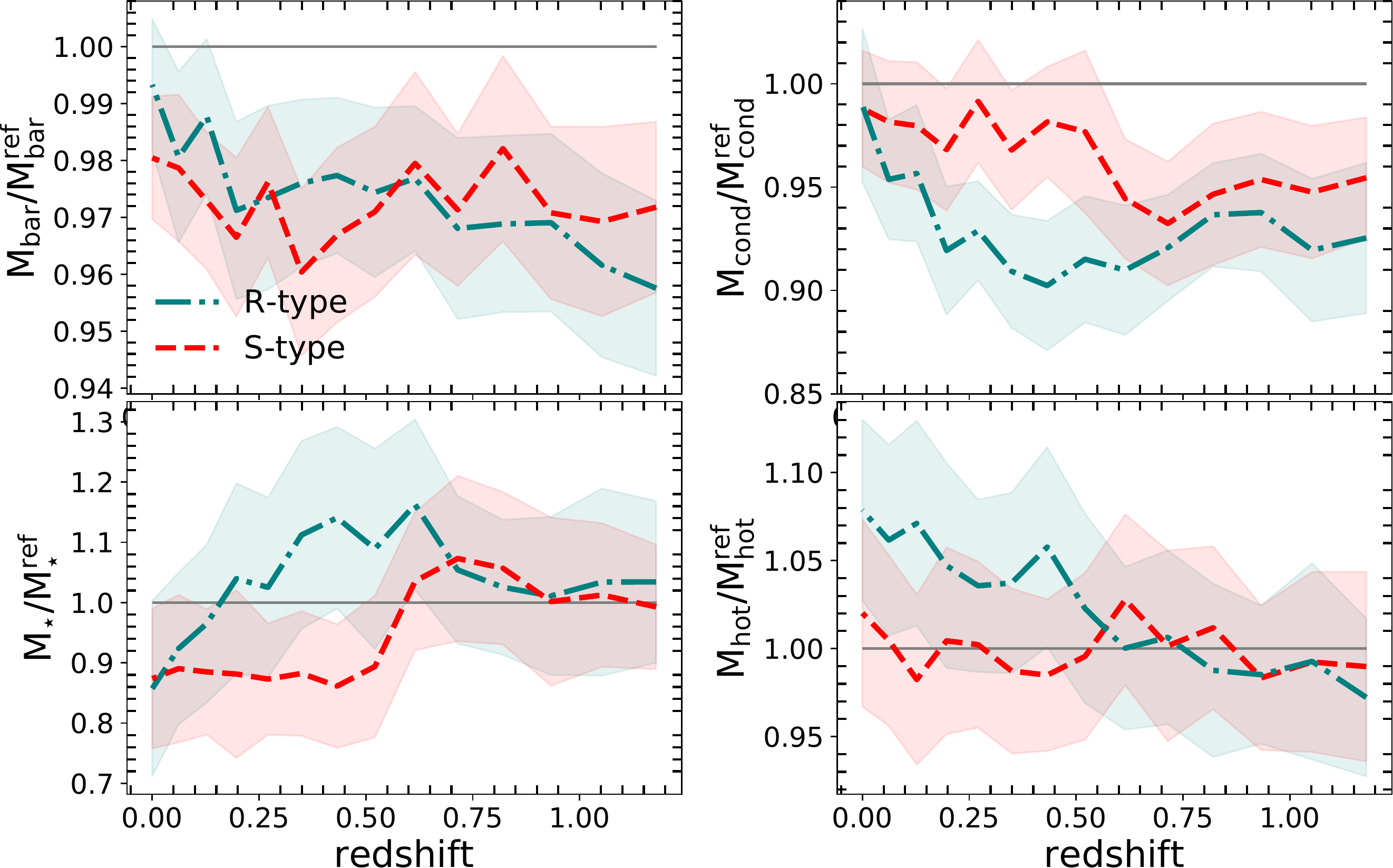}
    \caption{ Evolution with redshift of baryons masses in different stages in void haloes relative to reference simulation. The masses are calculated over the main branch of halo merger trees in voids in the mass range of $2\times10^{11}$ \msolh $<\rm M_{vir}<8\times10^{11}$\msolh at $z=0$. 
    The reference sample is constructed by randomly mimicking the mass distribution of haloes in voids at all snapshots. The shaded region corresponds to the error propagation of the mean.  
    The blue dot-dashed line is the region of R-type voids and the red dotted line is the region of S-type voids. The top-left panel shows the baryonic mass evolution, in the bottom-left panel the stellar masses, in the top-right panel the condensed masses and the bottom-right panel the hot masses. The horizontal grey line indicates the fraction equal to 1, corresponding to the reference sample. }
    \label{trees2}
\end{figure*}

It is important to note that the fractions of the different baryon phases are sensitive to the mass of the halo
\citep{Yoshida2002, Schaller2015}.
The differences in the mass evolution showed in Fig.~\ref{trees} may have an impact on the baryonic properties of these haloes as a function of the redshift. 
We are now interested in explore possible dependencies in the evolution of baryon properties of haloes with the type of void they inhabit.
To do so, we compare the evolution of the mass of each baryon stage for haloes in voids, with haloes in the reference simulation selected to mimic the mass distribution in voids at each redshift. 
In this way, each sample of reference to a given redshfit is independent of the other. 
The important fact is that the reference by definition perfectly follows the evolution of the halo mass in voids. As before, we obtain $100$ mass distributions from each snapshot in the reference simulation, with the same number than in the halo sample, and therefore the same statistical significant level. We use these samples to calculate the mean behaviour of each measurement and its error $\epsilon$.
In order to make a clean comparison between the two void types, we calculate the quotient between the void halo masses at different baryon stages and their corresponding comparison sample. 
Fig.~\ref{trees2} shows the evolution of the quotient of the fraction of baryons masses in the different stages in haloes in voids and in the reference sample. 
The top-left panel shows the halo baryon mass defined as the sum of stars and gas-particle masses inside $r_{\rm vir}$. Similarly, in the rest of the panels of this figure, we compute the mass of stars (bottom left), condensed gas (top right) and hot gas (bottom right) by counting particles inside $r_{\rm vir}$. 
In all the panels, the grey horizontal line indicates the fraction equal to 1, corresponding to the reference sample. As in Fig.~\ref{trees}, blue dot-dashed (red dashed) lines indicate the R(S)-type void haloes. The shaded area represents the error $\epsilon'$ calculated by propagating the error of the mean $\epsilon$ to this quotient. Recall that $r_{\rm vir}$ is defined at section \ref{subsec:cosmosim} as the radius at $200$ times the critical density of the Universe.
The haloes in both type of voids, show a lower baryon content compared to those selected from the reference sample. These haloes are around $2\%$ less massive with $\sim 1.5\epsilon'$ of significance level, independent of the void type. 
At the bottom-left panel, we see at $z=0$ the haloes belonging to voids have less stellar mass content ($10\%$ with $\sim \epsilon'$ of significance) respect to the reference simulation, but the halo content evolution shows dependencies with the void type. At $z=0.5$, haloes in R-type voids have more stellar-mass in comparison to the reference sample, $10\%$ above the reference with a $\sim\epsilon'$ of significance. On the contrary, at the same redshift S-type void haloes have a smaller stellar content respect to the reference halo, $10\%$ below the reference with a $\sim\epsilon'$ of significance. At higher redshift no significant differences can be seen.
At the top-right panel can be seen that the condensed gas content in voids is small respect to the reference sample. This is more strong in R-type void haloes, where the departures are close to $10\%$ below the reference with $\sim 3\,\epsilon'$ at $z\sim0.5$. 
The trend in the hot mass (bottom-right panel) is that R-type voids have $5\%$ more hot gas in comparison to the reference sample at $z=0$ (with $\sim \epsilon'$ of significance), whereas S-type voids haloes do not differ significantly from the reference. 

\section{Summary and conclusion}
\label{sec:summary}

In this work, we perform and analyse a set of hydrodynamical zoom-in simulations of cosmic voids and a cosmological box. We use these simulations to study the baryon differences in the void-environment in comparison with the general Universe. 
We analyse the dark matter haloes and their baryonic properties to understand how these particular environments impact the formation of galaxies.

We separate the void sample into two types depending on their dynamical state: R-type voids, which are expanding at all scales reaching asymptotically the mean density of the universe, and S-type voids, which are expanding in their inner regions but over compensated at large scales, therefore in contraction there.
Each void type is linked differently to the large-scale environment, with those in contraction having larger structures surrounding them.

We classify the baryon matter accordingly to five stages: diffuse gas, condense gas, hot gas, WHIM gas and stars.
Firstly, we study the density and velocity profiles of the dark matter and the different baryon stages for each resimulation and averaging over the different void types. As a first result, we see that the baryons closely follow the dynamics of the dark matter, no appreciable differences are seen in density and velocity profiles, which is consistent with \citet{Paillas2017}. 
These profiles are consistent with the expected dynamical behaviour of expansion and contraction of the void region depending on their void type.

In the inner void regions, the baryon content does not depend on the void type. These regions have a diffuse gas fraction twice of that the Universe in general. This overabundance translates into an underabundance of the processed gas phases, such as hot and WHIM. These differences are explained by the scarcity of large haloes and systems that process the gas through accretion shocks and star formation feedback. 
The regions near to voids also show differences in their gas content related to the presence of structures depending on the void type. Regions near S-type voids have abundant processed gas and little diffuse gas. The surroundings of this type of voids have abundant massive haloes, as clusters and FVS (Future Virialized Structures) \citep{Ceccarelli2013, Nadathur2015, Lares2017}. These massive haloes process the gas generating the observed differences. In contrast, the environments of the R-type voids are underdense regions, resulting in a lack of processed gas and abundant diffuse gas. 

We studied the phase-space diagram of matter surrounding haloes in voids, depending on the void type. In order to visualise the different components of matter infall, we defined mirrored phase space diagrams by subtracting at a given distance from the halo center the number of outflowing particles to the number of infalling particles. In this way, our analysis distinguished two infall components: the proper infall component related with particles at intermediate distances with moderate velocities, and the large scale flow component related with the movement of the halo onto larger-scale structures. 
We find that the dynamical state of dark matter and baryons around haloes in the inner regions of voids is very different respect to other environments.
However more novel results are found when this signal is analysed according to the type void, that is the void dynamics at very large scale. 
Haloes inside of R-type voids have a stronger proper infall component than haloes in S-type voids. 
The environment in the regions near voids also appears to affect the dynamics of gas around the haloes.
Haloes near to R-type voids have a large proper infall component of gas, meanwhile, haloes in near regions of S-type voids are mostly perturbed by the large scale flow. In this way qualitative similar trends are seen in phase diagrams of dark matter and gas around haloes.

The features in the phase-space diagram described above correlate with the matter density field surrounding the haloes. The two halo term of the halo-mass correlation function shows that the environment of haloes inside R-type voids is marginally less dense than S-type voids haloes. 
Also, the near void regions are different. In the near region of R-type voids the haloes are less clustered respecting to haloes in the near to S-type void region.
These features help us to interpret the differences in the phase-space results between R and S type voids. 
In particular, when we observe a strong correlation in the two-halo term for the near S-type void region, we find a strong signal in the large-scale flow in the phase-space diagrams. 

All these results are somewhat consistent with those obtained by \citet{Ruiz2015}, where they study the inner structure of voids in dark matter only simulation accordingly to void dynamical state. Their results on the distribution of velocities within the void radius indicate that the non-linear dynamics of the inner regions of voids depend on the void type. They found that the velocity field in S-type voids have a broader distribution indicating that haloes are infalling into larger structures in a consistent way with our results.

We constructed the merger tree and found that haloes in the voids grow more slowly compared to the general population of haloes in the Universe. This trend is particularly strong in haloes belonging to R-type voids. We found that the halos in these voids have systematically significant smaller masses than in the reference within redshift between $0$ and $1.2$.
This is consistent with results by \citet{Alfaro2020} using the same void definition as in this work, as well as employing different void definitions \citep[][ among others]{Tonnensen2015,Tojeiro2017,Martizzi2020}.
This seems to indicate that filamentary structure present in denser environments can lead to a more efficient halo formation process 
than in an underdense environments. 
Our contribution to this discussion is that the assemble of haloes seems to depend also on the large scale environment. R-type void haloes assemble later than S-type void haloes, while the density in their inner regions is almost identical.

We find at $z=0$ and at a fixed halo mass, void haloes have less stellar content than haloes in the general Universe. This has been pointed out for both semi-analytical models and hydrodynamical simulations in recent papers \citep{Tonnensen2015,Tojeiro2017,Xu2020,Alfaro2020,Habouzit2020,Martizzi2020}. 
The comparison of our result with \citet{Alfaro2020} is direct. We find consistently, that there is a $\sim 10\%$ less stellar mass in void haloes at a given mass range concerning to the general halo population.
From the observational side, \citet{Florez2021} report that void-galaxies have a higher gas-to-stellar-mass ratio than galaxies in more dense environments. 
This can be interpreted as a consequence that void-galaxies have less stellar content at a given halo mass.

At high redshift ($z>0.75$) the stellar content in void haloes and the reference sample are similar.
There is a margin trend at $z\sim0.4$ to have higher stellar masses in R-type void haloes compared with the S-type, showing some dependence on the star formation efficiency with the redshift and the environment. 
When we analyse the baryon mass evolution, we find a general trend of lower baryon fraction in haloes for void environments \citep[consistently with][]{Xu2020}.

The condensed and hot gas fractions also differs in void haloes and the reference sample. 
The content of condensed gas is lower in void haloes in agreement with the lower baryon fraction. 
Moreover, we have a lower fraction of condensed gas in the R-type voids at the same epoch when there is an increment in the stellar mass fraction due to the star formation process. 
In this way, the condensed gas is depleted by the star formation process while the associated feedback mechanisms suppress the condensation of the gas. 
The hot gas fraction is marginally higher in S-type and larger in R-type voids.
The hot gas can form by accretion shocks which are related to isotropic infall. This combined with the fact that R-type void haloes assemble later than S-types, suggest that they have fewer cosmological structures (i.e. filaments) that help to smoothly accrete material into haloes.

Finally, we want to highlight the fact that the largest differences in condensed and hot gas content, stellar masses and assembly time are noted in the R-type voids, even with the limitations in the number of re-simulations and resolutions used in this work.
As we mentioned before, the relative fraction between R and S - type voids is dependent on the void size \citep[see for instance ][]{Ceccarelli2013}. S-type voids dominate the statistics of smaller voids whereas R-type voids are more predominant at large void sizes. The present work is restricted to voids in a very narrow range in void size, however we could be tempted to extrapolate our results as a prediction of the mean behaviour of voids as a function of their size. In order to do this it should be assumed that the R-type and S-type voids analysed in this work are representative of they void type at all sizes. However, to check this assertion larger void simulations should be performed.
Acknowledged this, the stellar mass fraction differences in redshift and the dependence with the environments show that we will be able to compare directly with the new generation of observational surveys, such as the Legacy Survey of Space and Time, LSST, at the Rubin Observatory \citep{Ivezic2019} and theoretical predictions from numerical simulations.  
In future works, we want to explore the use of different star formation recipes showing that void regions are useful environments to constrain star formation models with observations.

\section*{Acknowledgements}
We thank the anonymous referee for the very helpful comments that greatly improved this paper.
This work was partially supported by the Consejo de Investigaciones Científicas y Técnicas de la República Argentina (CONICET) and the Secretaría de Ciencia y Técnica de la Universidad Nacional de Córdoba (SeCyT).
ARM is doctoral fellow of CONICET. DJP, FAS and ANR are members of the Carrera del Investigador Científico (CONICET).
ARM, FAS and ANR thank support by grants PIP 11220130100365CO, PICT-2016-4174, PICT-2016-1975 and Consolidar-2018-2020, from CONICET and FONCyT (Argentina); and by SECyT-UNC.
This work would not have been possible without access to the computational resources of the Centro de Computación de Alto Desmpeño of the Universidad Nacional de Córdoba (CCAD-UNC) which is part of SNCAD-MinCyT, Argentina.

\section*{Data Availability}

Data is available on request to the corresponding author. 

\bibliographystyle{mnras}

\input{paper.bbl}

% Don't change these lines
\bsp	% typesetting comment
\label{lastpage}
\end{document}